\documentclass[final,5p,times,twocolumn]{elsarticle}
\usepackage{amssymb}

\date{\today}

\usepackage{color}

\begin{document}
\begin{frontmatter}
\title{ Role  of the cluster structure of $^7$Li in the dynamics of fragment
capture}

\author[barc]{A. Shrivastava \corref{cor1}}
\cortext[cor1]{Corresponding author}
\ead{aradhana@barc.gov.in}

\address[barc]{Nuclear Physics Division, Bhabha Atomic Research Centre, Mumbai
400085, India}

\author[ganil]{A. Navin}
\address[ganil]{GANIL, CEA/DSM - CNRS/IN2P3, Bd Henri Becquerel, BP 55027,
F-14076 Caen Cedex 5, France}

\author[ECT]{A. Diaz-Torres}
\address[ECT]{ECT$^*$, Villa Tambosi, I-38123 Villazzano, Trento, Italy}

\author[TIFR]{V. Nanal}
\address[TIFR] {DNAP, Tata Institute of Fundamental Research, Mumbai 400005,
India}

\author[barc]{K.~Ramachandran}
\author[ganil]{M.~Rejmund}
\author[vecc]{S.~Bhattacharyya}
\address[vecc]{Variable Energy Cyclotron Centre, 1/AF Bidhan Nagar, Kolkata
700064, India}
\author[barc]{A.~Chatterjee}
\author[barc]{S.~Kailas}
\author[ganil]{A. Lemasson}
\author[TIFR]{R.~Palit}
\author[barc]{V.V.~Parkar}
\author[TIFR]{R.G.~Pillay}
\author[barc]{P.C.~Rout}
\author[TIFR]{Y.~Sawant}

\begin{abstract}
    Exclusive measurements of prompt $\gamma$-rays from the heavy-residues with
various  light charged particles in the $^7$Li +  $^{198}$Pt  system,  at an 
energy  near  the  Coulomb  barrier (E/$V_b$~$\sim$~1.6) are  reported. Recent
dynamic classical trajectory calculations, constrained by the measured fusion,
$\alpha$  and $t$   capture  cross-sections have been used to explain the
excitation energy dependence of the residue cross-sections. These calculations
distinctly illustrate a  two step process,    breakup followed by fusion in case 
of the capture of $t$  and $\alpha$ clusters; whereas for $^{6}$He~+~$p$ and
$^{5}$He~+~$d$ configurations, massive transfer  is inferred  to be the dominant
 mechanism.  The  present work  clearly  demonstrates the role played  by the
cluster  structures of $^7$Li  in understanding the reaction dynamics at
energies around the Coulomb barrier.
 \end{abstract}


\end{frontmatter}

   In  weakly bound  nuclear systems,  correlation among  nucleons and pairing 
are manifested,  among others,  as an  emergence  of strong clustering  and
exotic  shapes. This  has renewed  interest  in the understanding of  clusters
based  on concepts of  molecular physics and     the     role    of     cluster 
states    in     nuclear synthesis~\cite{his12,freer07a}.   Lithium isotopes
present a unique  example  of   nuclear  clustering,  with  lighter  isotopes
($^{6,7}$Li) having  a well known $\alpha$~+~$x$  cluster structure and the 
heaviest bound  isotope ($^{11}$Li)  exhibiting a  two neutron Borromean
structure.   $^9$Li has  also been described  as $^{6}$He +~$t$  in a  recent 
work~\cite{arxive}. $^7$Li  is an  equally interesting  case with  its well 
known weakly  bound $\alpha$~+~$t$ structure (S$_{\alpha/t}$ = 2.47 MeV), as
well as less  studied more strongly bound clusters $^6$He~+~$p$~(S$_{^{6}He /p}$
= 9.98 MeV) and $^5$He~+~$d$ (S$_{^{5}He/d}$ = 9.52 MeV)~\cite{Til02,li10}. 

  Recent studies  with weakly  bound nuclei have  also focused  on the
understanding  of  the role  of  novel  structures  in the  reaction
dynamics~\cite{nick}.   Dominant reaction modes  in nuclei  with low binding
energies,  involve inelastic excitation to  low lying states in the continuum or
transfer/capture of one of the cluster fragments from  their   bound/unbound 
states  to   the  colliding  partner nucleus~\cite{nick,alexis,alexis2}. The
role of inelastic excitation of low lying  unbound  states and transfer in the 
fusion hindrance, observed  at  energies well below the  barrier, is also  a
topic of current interest~\cite{Jiang06,ara09}.  When the capture occurs from
unbound states of  the projectile, the process could  be looked upon as a  two
step process,    breakup followed by  fusion (\emph{breakup
fusion})~\cite{cast78,utsu83,vanbrief}.   In  case  of well  bound nuclei,
nuclear  reaction related to  capture of heavy  fragments by the  target has
been  identified as  incomplete  fusion or  massive transfer~\cite{pushp}  and
occurs  predominately  at energies  $\ge$ 10~MeV/A.  For  weakly bound cluster
nuclei such  as $^{6,7}$Li, the former has  been shown  to be important  both
above and  at energies much  below  the   Coulomb  barrier \cite{ara09,russian}.
  Earlier studies have found the process of breakup fusion to be more dominant
over a one step transfer  in case of $^{6}$Li($^{7}$Li) for deuteron (triton)
capture reaction ~\cite{cast78,utsu83,vanbrief}.

A number of theoretical approaches for understanding the incomplete fusion process have been developed, as recently reviewed in \cite{thomp,psingh,mier12}. These span a range of concepts and considerations, including breakup-fusion,  angular momentum window for incomplete fusion, promptly emitted particles,  Fermi-jet,   exciton, and   moving source, thereby  explaining the measured energy spectra and angular distribution of the emitted fragments and  population of angular momentum in the compound system.
  Recently a theoretical description of  breakup fusion for weakly bound nuclei
has been incorporated  in the three-dimensional classical trajectory
model~\cite{alexis},  considering  the  peripheral nature  of  the  process  for
predicting   both  the  spin distribution and sharing of excitation energy.  The
 model treats the breakup process stochastically and follows  the time 
evolution of  individual fragments  for  a unique identification of reaction
processes.  More recently it has been  improved to  include  the time 
propagation  of the  surviving breakup  fragment  and  the  breakup fusion 
product,  allowing  the description of their asymptotic angular distribution
\cite{alexis2}. In contrast to most existing models for incomplete fusion, the new approach~\cite{alexis,alexis2} 
treats the dynamics of incomplete fusion and provides a number of differential cross sections that are critical for 
understanding exclusive experimental data. Such a theoretical model now allows for the first time to interpret
data from exclusive experiments, which was not possible earlier.

The  present  work is  aimed  at exploiting  the  above  model \cite{alexis,alexis2} to
understand  the dynamics of  the process of fragment capture for
the   various  cluster   structures ($\alpha$~+~$t$,   $^{6}$He+$p$  and
$^{5}$He+$d$) of $^7$Li, using exclusive particle-gamma coincidences to
uniquely  identify   the  source  of  the   various  residues  formed.
Integrated  cross-sections  of   compound  nuclear  fusion,  $t$  and
$\alpha$-capture  using both  off- and  in-beam gamma  decay  along with
yields of  the evaporation residues for  different excitation 
energies of  the composite  system  are  reported.  These results  are
compared  with  those of the   recent  three-dimensional  classical  trajectory
model~\cite{alexis,alexis2} in conjunction with the statistical model of compound nucleus evaporation.  
 This  comparison  demonstrates for  the first  time the  important role  played  by the  cluster structures  of
$^7$Li in interpreting the dynamics  of fragment capture.

Two independent experiments were performed at the 14UD pelletron
Facility-Mumbai:   a)  Measurement  of  the  prompt $\gamma$-rays  from the
heavy    residues in coincidence with  various light particles  $\alpha$, $t$,
$d$ and $p$.   The  $\alpha$-capture cross-sections were measured employing
in-beam gamma method b)   Measurement  of   the excitation function  for fusion,
$t$  -capture and  neutron  transfer  using radioactive decay  of the residues.

 The  measurements for exclusive in-beam $\gamma$ decay of the residues were
performed  using  a  $^7$Li  beam  of energy  45~MeV,  incident  on  a
1.3~mg/cm$^2$ thick  self supporting foil of $^{198}$Pt  with a 95.7\%
enrichment.     Four   telescopes    ($\Delta$E$\sim$25-30$\mu$m   and
E$\sim$1mm)   at   50$^{\circ}$,   60$^{\circ}$,   120$^{\circ}$   and
130$^{\circ}$  (covering the  region near  and away  from  the grazing angle)
were  used to measure  the charged particles.   Four efficiency calibrated 
Compton  suppressed   clover  detectors,  operated  in  an add-back mode,  were
placed at  14.3 cm from  the target at  angles of 35$^{\circ}$,  -55$^{\circ}$, 
 80$^{\circ}$,  and  155$^{\circ}$.   A coincidence between any charged particle
recorded in the $\Delta$E and a $\gamma$-ray   in  any  clover  detector  or  a 
two fold  $\gamma$-ray coincidence between  clover detectors was used as  the
master trigger. The reaction products arising  from different channels were
identified by their  characteristic $\gamma$-ray transitions  in coincidence
with the outgoing particles.  In  the following we discuss the $\gamma$-ray
spectra  obtained by  selecting  different ejectiles  recorded in  the
telescopes placed  at 50$^{\circ}$ and 60$^{\circ}$  that cover region around
the  grazing angle.  At backward angles,  the contribution from fragment capture
reaction was verified to be negligible.
\begin{figure}[h]
\includegraphics[width=20pc]{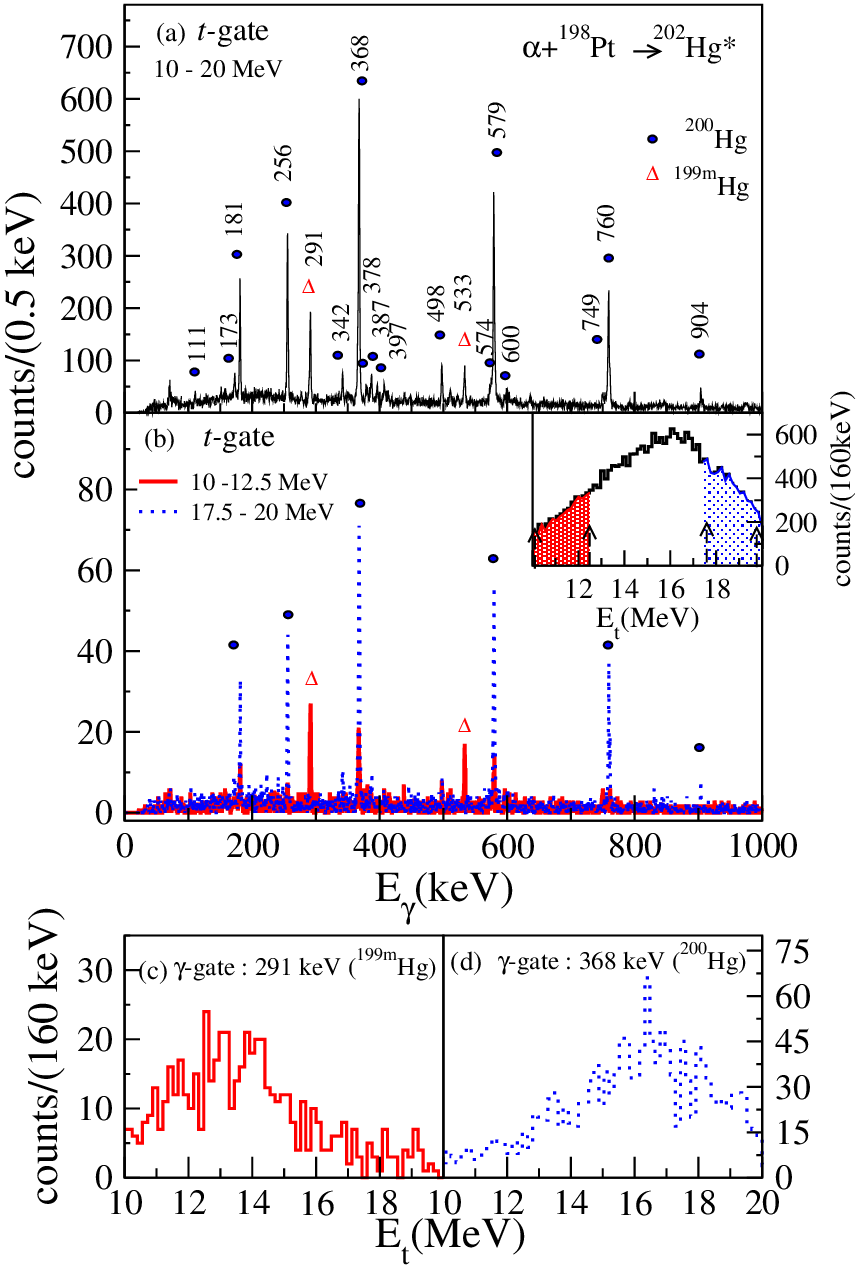}
\caption{(color online) 
  (a)  Prompt  $\gamma$-ray   spectra  obtained  in  coincidence  with
  outgoing $t$  ($\alpha$-capture) having an  energy 10 to 20  MeV (b)
  same as (a) but subdivided into two different energy bins of the $t$
  (10 to 12.5  MeV and 17.5 to 20  MeV as shown in the  inset) (c) and
  (d)  Triton energy  spectrum  at 50$^{\circ}$,  in coincidence  with
  291~keV  and  368~keV  $\gamma$-rays  arising  from  transitions  in
  $^{199}$Hg and $^{200}$Hg respectively}
\label{fig:fig1}
\end{figure}

  Plotted  in  Fig.~1a  is  the  $\gamma$-ray spectrum  gated  by  the outgoing
tritons with  kinetic energy between 10 to  20 MeV, showing  peaks from 
the residues of the  composite system, $^{202}$Hg, corresponding  to capture  of
$\alpha$-particles  by  the $^{198}$Pt target.    The  $\gamma$-ray transitions
from  yrast-bands of $^{200}$Hg,   as  reported  in  Ref.~\cite{help81}  for
$\alpha$+$^{198}$Pt system, are  observed.  In case of $^{199}$Hg, the
$\gamma$-ray  transitions  feeding  the   long  lived  isomeric  state 
(13/2$^+$, T$_{1/2} \sim$  42.8~min), known from  an earlier study  in $\alpha$
+$^{198}$Pt system~\cite {mert78}  are labeled.  The triton spectrum from the 
two telescopes  was further divided  in to  smaller energy bins (2.5 MeV) to
study  the variation in population of the residues as  function  of  triton 
energy   and  is  shown  in  Fig.~1b.  The $\gamma$-ray spectra  related to the 
two bins are plotted  as solid and dotted curves. The  outgoing tritons with
higher (lower) kinetic energy correspond  to lower (higher) excitation  energy
deposited in $^{202}$Hg.  As can  be seen from the figure  the $\gamma$- peaks of
the residues  corresponding  to  the  two  neutron  evaporation  channel
($^{200}$Hg) are  dominant with the  higher triton energy  bin while those  from
 three  neutron  evaporation  channel  ($^{199}$Hg)  are dominant with  the
lower  energy bin.  A  similar conclusion  can be obtained  from  Figs.~1c,d 
that  show triton  spectra  obtained  by selecting  the 291  keV and  368 keV 
$\gamma$-ray  transitions from $^{199}$Hg and $^{200}$Hg  respectively.  The
relative population of $^{200}$Hg  corresponding  to  each  bin  of  the  triton
 spectrum were  estimated  from   the  efficiency  corrected   yields  of 
$\gamma$-ray transition  to the  ground  state.  A similar procedure  was
followed  for $^{199}$Hg for transitions above  the isomeric  state, 13/2$^+$ at
532 keV.

\begin{figure}[h]
\includegraphics[width=20pc]{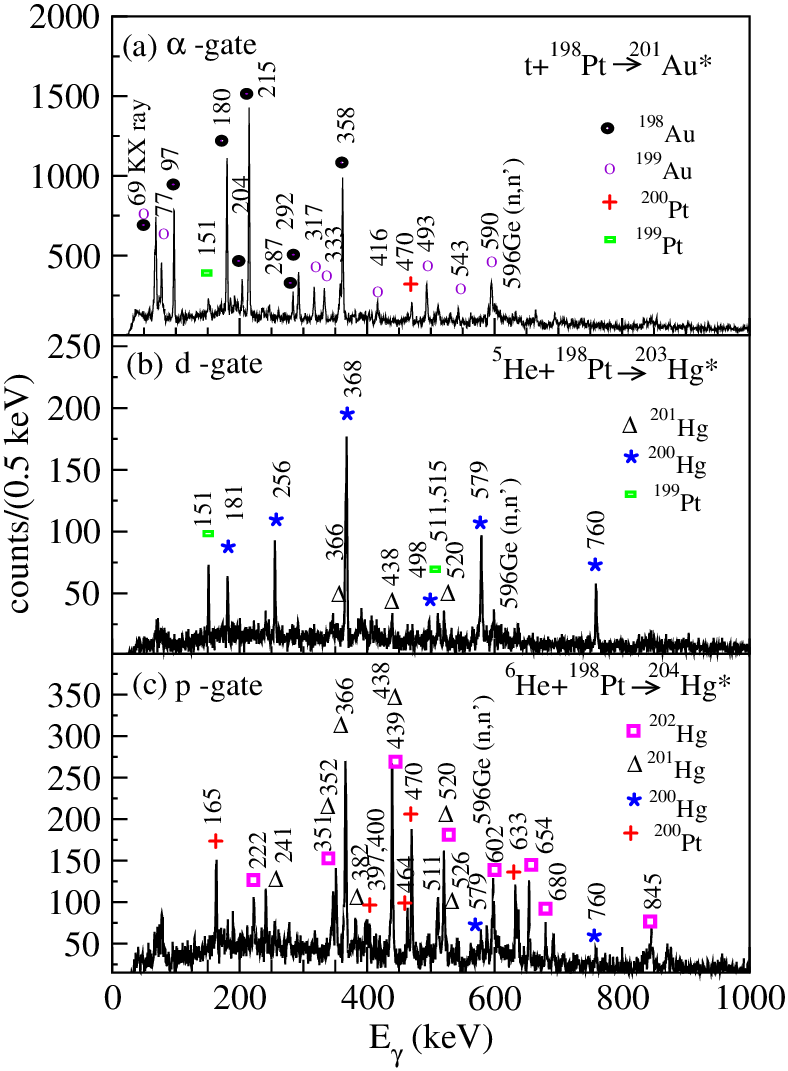}
\caption{(color online)   
In beam  $\gamma$-ray spectra  in coincidence with  outgoing fragments
(a)  $\alpha$, (b) $d$  and (c)  $p$. The  gamma transitions  from the
residues  populated  from capture  of  $t$,  $^5$He and  $^6$He  are
indicated in  (a)-(c) respectively. The $\gamma$-rays  arising from 1n
and 2n-transfer ($^{199,200}$Pt) are also labeled. }
\label{fig:fig2}
\end{figure}

 In Fig. 2a,  the $\gamma$-ray spectrum obtained in  coincidence with
$\alpha$-particles  shows   contribution  arising  from   different  reaction
  channels. The  main $\gamma$-ray transitions in the  spectrum arise from
  $^{198,199}$Au~\cite{mahn75} (residues due to $t$-capture).
    Shown in  Figs.~2b,c are the $\gamma$-ray spectrum  in coincidence with the
deuterons  and protons respectively.   Comparing these spectra with Fig.~1a,  it
can be  noticed that more neutron  rich residues (due  to  capture  of  the
heavier  complementary  particle),  are populated in  going from spectra  in
coincidence with $t$  to $p$. In the  $\gamma$-ray  spectrum gated  by  the
deuterons,  the  peaks arise mainly from  the residues that can be
attributed to  decay of     $^{203}$Hg  arising from  the  capture of  $^5$He
(Fig~2b).  The  known $\gamma$-ray transitions  from $^{200,201}$Hg could be
identified. The dominant peaks  observed in Fig.~2c are from the  $^{201,202}$Hg
residues, of  the composite  system $^{204}$Hg formed after capture of $^6$He.

The $\gamma$-ray transitions from $^{199,200}$Pt, corresponding to one and two 
neutron transfer  reactions are observed  in the $\alpha$, $d$  and $p$ gated 
spectra.  This  can be  understood as  follows. The ejectile  resulting from one
 neutron transfer  ($^6$Li) has  no bound excited  state.  The  excited  states
populated  in  the continuum  of $^6$Li    after   transfer   of    a   neutron,
   disassociate   into $\alpha$~+~$d$~\cite{ara06}.   These events  are 
identified from  the characteristic  $\gamma$-ray  transitions  of $^{199}$Pt 
observed  in coincidence   with  $d$   (Fig.~2b).   Spectroscopy   information 
for $^{199}$Pt is  presently limited.  The unidentified   peak at 151 keV  
has  been   tentatively  assigned   to  $^{199}$Pt   since  this $\gamma$-ray
transition was also observed  in coincidence with $^6$Li. The $\gamma$-ray
transitions from $^{200}$Pt  observed in coincidence with $p$ in Fig.~2c are
attributed to  the breakup of excited unbound states in $^5$Li  into 
$\alpha$~+~p after  transfer  of  two  neutrons. In  the $\alpha$-gated 
spectrum  (Fig.  2a),   peaks  arising  from $^{199,200}$Pt are small
compared to dominant  peaks arising from residues  from $t$-capture,  which
are  formed with  relatively larger cross-sections. This was confirmed from the
offline- measurement of the  neutron transfer and $t$-capture cross-sections (as
shown in Fig.~3 and discussed below).

\begin{figure}
 
 \includegraphics[width=18pc]{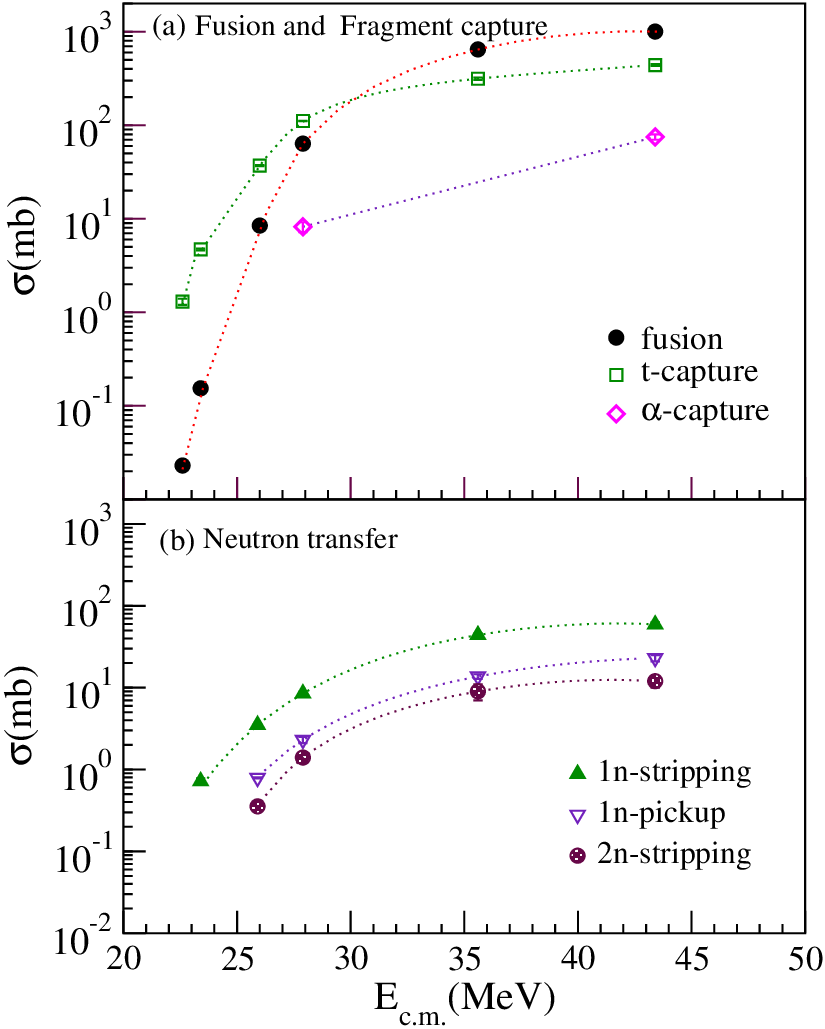}  
\caption{(color   on line) Integrated cross-sections for (a) compound-nuclear
fusion ($^{199-202}$Tl), $t$-capture ($^{198-200}$Au) and $\alpha$ -capture
($^{199-200}$Hg)  (b) neutron transfer corresponding to  1n-pickup  ($^{197}$Pt)
and  1n, 2n-stripping  ($^{199,200}$Pt)  reactions.   The  dashed  lines in
panels (a) and (b) are  to     guide the eye.   }
\label{fig:fig3}
\end{figure}

The   peaks  at 366~keV  and 241~keV  in Fig.~2c  could  not be identified 
among the  known transitions  of  $^{200,201,202}$Hg and $^{200}$Pt. A  probable
candidate could be from the decay of states above   the  13/2$^+$   isomer  in  
$^{201}$Hg.    No  spectroscopy information  of  the prompt gamma  transition 
above  this state  is presently available in literature.  Change in $\gamma$-ray
intensity of these transitions as  compared to transitions from $^{202}$Hg was
studied with different energy  bins of the scattered proton, further confirming 
this assignment to  $^{201}$Hg.  This  observation shows advantage of the
breakup fusion reaction for studying nuclear states at   higher   spin,  not  
accessible   by   the  compound   nuclear fusion, earlier demonstrated in
Ref.~\cite{anu}.

 The cross-section  of the residues, $^{199}$Hg  and  $^{200}$Hg from
$\alpha$-capture were  deduced  by performing  coincidence between  $\gamma$-ray
 transitions from  the  yrast  band built  on ground  state  in  case  of 
$^{200}$Hg and  on  the  isomeric  state (13/2$^+$) in case of  $^{199m}$Hg. The
coincidence condition improved the quality of the spectra  especially for the
peaks that could not be resolved from the peak of interest. The
cross-section for $^{199m}$Hg was also obtained by following the radioactive
decay of the isomeric state (13/2$^+$). This was found to be in a good agreement
 with the value extracted from the in-beam method.    The   $\alpha$-capture 
cross-sections,   obtained by  adding cross-section of  $^{199m}$Hg  and that of
$^{200}$Hg are shown in Fig.~3a at beam energies of 29 and 45 MeV.

 The measurement  of the cross-section of the  residues resulting from the
process  of fusion, $t$-capture and  neutron transfer channels  were performed
with beams of $^{7}$Li in the range  of 22 to 45 MeV incident on self supporting
 foils  of $^{198}$Pt  target  (95.7$\%$ enriched,  $\sim$ 1.3~mg/cm$^2$  thick)
followed  by an  Al catcher  foil  of thickness $\sim$ 1~mg/cm$^2$.  Two
efficiency calibrated HPGe detectors with Be window were used in a low background
counting setup with graded shielding for off-beam gamma ray measurements. The 
residues in  case of  fusion ($^{199-202}$Tl) were  identified  by using 
KX-$\gamma$-ray  coincidence of  the decay  radiations from  the irradiated 
sample with  detectors placed face  to  face.  Further  details  of  the  setup 
can  be  found  in Refs. \cite{ara09,fusion11}.

The  $\gamma$-ray   yields  for  residues   formed  after  $t$-capture
($^{198-200}$Au)  and for target-like nuclei after neutron transfer
($^{197,199,200}$Pt) were    extracted   from   inclusive   $\gamma$-ray
measurements.  The $t$-capture and fusion cross-sections were obtained by 
taking   the  sum  of  individual   measured  evaporation  residue
cross-sections. The  excitation function for fusion, fragment capture and
neutron transfer reactions are plotted in Figs.~3a,b; respectively. Corrections
for $^{196}$Pt impurity  in the target (2.56\%) were found to be  negligible ($<
1\%$).   The error on the  cross-section arises mainly from statistics.  Other
sources  of error in the cross-sections, arising from  the uncertainties in 
measurements of the  beam current, $\gamma$  -ray  efficiency,  target 
thickness and available spectroscopic information  of  the residues  were 
estimated  to  be between  10  to 15\%. The cross-sections for fusion and fragment capture at the highest energy (45 MeV) are also listed in Table.1.

\begin{table}[htb]
\begin{center}
\caption{ {\sc platypus} calculation for  integrated $\alpha$-capture, 
$t$-capture and complete fusion cross-sections in $^7$Li +
$^{198}$Pt~\cite{fusion11} 
at a beam energy of 45 MeV.  }
 \begin{tabular}{@{}lllll}
\hline
 reaction channel &  $\sigma$  &  $\sigma$   
   \\
  &{expt}& { {\sc platypus} } \\
   & (mb) & (mb)   \\
\hline
 $\alpha$ -capture &  75~$\pm$5   & 77    \\
 $t$ -capture &  440~$\pm$10   &   366     \\
  complete fusion &  1004~$\pm$50  &  958   \\

\hline 
\end{tabular}
\label{tab1}
\end{center}

\end{table}

   A   three-body   classical  dynamical   model~\cite{alexis,alexis2}
implemented in  the {\sc  platypus} code~\cite{platypus}  allows a consistent 
analysis of  breakup, incomplete,  and  complete fusion processes.   The 
measured  integrated cross-sections  of  complete fusion,    $\alpha$-capture  
and    $t$-capture   for    the $^7$Li~+~$^{198}$Pt  system at 45~MeV (Fig. 3a and
  Table~1)   were used  to  constrain the parameters   of  the  model   while 
predicting   the  differential cross-section as a function of excitation energy
and angular momentum for the formation of the primary composite system. 
These differential cross-sections are the key inputs to the  statistical model  code~{\sc
pace2}~\cite{gav80} for calculating the cross-section of evaporation residues from decay of the composite system.

  The {\sc platypus}  calculations were carried out  considering $^7$Li as
$\alpha$ + $t$ cluster, having a binding energy of 2.47 MeV. In this calculation
breakup  fusion occurs when any of  the breakup fragment ($\alpha$  or  $t$) 
penetrates  the  Coulomb  barrier  between  the fragment  and the  target.
Complete  fusion occurs  when  the entire projectile, $^7$Li or both $\alpha$ 
and $t$ get captured inside the interaction  barriers. Parametrization  of the 
Coulomb  and nuclear potential  was    same  as in  Ref.~\cite{alexis}.   The
pre-  and post-breakup   Coulomb   and   nuclear  interactions   between   the
participants         ($^7$Li+$^{198}$Pt,        $\alpha$+$^{198}$Pt,
$t$+$^{198}$Pt,        $t$+$\alpha$,        $t$+$^{202}$Hg       and
$\alpha$+$^{201}$Au), were taken as those between a point charge and a spherical
distribution for the heaviest fragment (= 1.2A$^{1/3}$). The nuclear 
interaction was parametrized by  a Woods-Saxon potential (Table~\ref{tab2})
which provided  Coulomb barriers similar to those of the Sao Paulo
potential~\cite{SP_pot1}.

Parameters   necessary  for   the  breakup-probability function  [$A \exp(-\beta
R)$, where $R$ denotes the internuclear distance] were obtained  by  reproducing
  the  measured  integrated cross-section of  $t$-capture and  $\alpha$- capture
and  the complete fusion for $^7$Li  + $^{198}$Pt (Table 1). The  amplitude and
slope of the break-up  probability function were $A=1.68 \times  10^4$ and
$\beta = 0.9$ fm$^ {-1}$,  respectively \cite{alexis2}.  Location of  the
breakup of $^7$Li into $\alpha$ and $t$ was determined by Monte Carlo sampling
of the  breakup  function  up  to  $R=50$ fm.   The  instantaneous  dynamical
variables of the  excited projectile at the breakup  point, namely its total 
internal energy  ($E_{rel} \leq 6$ MeV),  its angular  momentum  (up to
4$\hbar$), the fragment  separation  and their orientations, were also Monte
Carlo sampled \cite{alexis2}. The cross-section for $\alpha$ and $t$  capture 
were  calculated   using  a  sharp  cut-off  in  angular momenta. The
convergence of the calculated cross-section was ensured by including
projectile-target partial waves up to 50$\hbar$, for a sample of 1000
projectiles  per partial wave. The calculations  were found to be  in agreement
with  the shape  of the  measured energy  spectrum of surviving 
$\alpha$-particles  and $t$.   As  the  fusion barrier  for $t+^{198}$Pt  is 
smaller  than  that for  $\alpha+^{198}$Pt,  the cross-section    for   $t$-   
capture   is    larger    relative   to $\alpha$-capture. The comparison between
the calculation and the measured values are shown in Table \ref{tab1}.

\begin{table}[htb]
\begin{center}
\caption{ Woods-Saxon potentials employed in the $^7$Li + $^{198}$Pt 
{\sc platypus} calculation. The strength $V_0$ is in MeV, while the 
radius $r_0$ and diffuseness $a_0$ parameters are in fm.}
 \begin{tabular}{@{}lllll}
\hline
 mass partitions  &  ($V_0$,~$r_0$,~$a_0$)  \\
\hline
  $^7$Li + $^{198}$Pt &  {(-41.58,~~ 1.640,~~ 0.630) } \\
  $\alpha$ + $^{198}$Pt & {(-10.26,~~ 1.534,~~ 0.587)} \\
  $t$ + $^{198}$Pt & {(-26.40,~~ 1.562,~~ 0.595)} \\
  $t$ + $\alpha$ & {(-18.16,~~ 1.284,~~ 0.630)} \\
  $t$ + $^{202}$Hg & {(-26.40,~~ 1.562,~~ 0.590)} \\
  $\alpha$ + $^{201}$Au & {(-10.26,~~ 1.534,~~ 0.584)} \\
\hline 
\end{tabular}
\label{tab2}
\end{center}

\end{table} 


 To  get  further  insight  into the  mechanism  of  fragment-capture, the
measured yields of the  evaporation residues obtained from the particle-gamma
coincidence data    were  compared with  the predictions from {\sc platypus + pace2}  
for different excitation energies  (E$^*$) of  the  primary composite system as
discussed below.  The spectrum of the surviving $\alpha$-particles,  after
capture of the complementary fragment ($t$), represents the  cross-section for
breakup-fusion as a function   of   the   kinetic   energy  of   the  
$\alpha$-particles (E$_\alpha$).  This can be expressed as a function of E$^*$ 
of the composite  system $^{201}$Au,  by obtaining  the E$^*$  for each value of
E$_\alpha$, using the  dynamical variables at the instant of breakup of $^7$Li
into $\alpha$+$t$  on an event by event basis.  The calculated E$^*$  and the
corresponding  breakup fusion cross-section as a  function of spin  ($\sigma_J$
vs J)  were given as input  to the statistical model  code~{\sc
pace2}~\cite{gav80} for  calculating the evaporation  residue cross-sections
from  decay of  $^{201}$Au formed after   triton-fusion.     The   calculated  
values    of   absolute cross-sections for the residues, $^{198,199}$Au, are
plotted as solid and dashed curves in Fig.~4a.  The measured yields of
$^{198}$Au from the second bin of (bin II in inset Fig. 4a)  $\alpha$-particle spectrum, were normalized
to the calculated cross-section  obtained using {\sc pace2} for  the E$^*$ = 30
MeV that corresponds to the E$_\alpha$ = 24 MeV (center of the bin used).  The
cross-section for  $^{198,199}$Au deduced  after applying the same normalization
to their respective yields in each bin (bin I, III and IV in inset of Fig. 4a) and are plotted in Fig.~4a. The errors on
cross-sections are only statistical in  nature.   A  reasonably  good  agreement
 is  observed  with  the calculation.    These  results  suggest   that  the  
main  mechanism responsible  for $t$- capture  is fusion  of $t$  after  breakup
of $^7$Li, as  modeled in the  {\sc platypus} code.  Following  the same
procedure, cross-sections  for residues  arising from the  capture of
$\alpha$-particles  for  a given  energy  (corresponding to  outgoing triton
energy,  inset of Fig.~4b)  were calculated from  {\sc pace2}, using
spin distribution and  excitation energy of $^{202}$Hg obtained from  {\sc 
platypus}  and  are  shown in  Fig.~4b.   The  calculated cross-section of
$^{200}$Hg at E$^*$=27 MeV was used to normalize the measured yield of
$^{200}$Hg and $^{199}$Hg. In case of  $^{199}$Hg the $\gamma$ -ray transitions
only above the (13/2$^+$) isomeric state  were considered hence the measured
cross-sections only provide a lower limit for this channel. The energy
dependence of formation of both the  residues agrees  well with the  statistical
 model calculations, showing a similar dominance of the breakup fusion process.
The {\sc platypus} calculations indicate that the breakup fusion process is
dominated by breakup events with $E_{rel} \leq 4$ MeV, which only includes
prompt breakup. This type of breakup is critical, as the resonant  states  have 
life   time  larger  than   the  interaction time~\cite{luo11}. It can be
noticed from the inset of Fig.~4a,b that both  $\alpha$ and  $t$ spectra  peak 
around the  beam velocity,  as expected from the breakup fusion model of {\sc
platypus}.

 A similar analysis  was attempted by modeling $^7$Li  as a cluster of
$^6$He+$p$ (breakup  threshold 9.975~MeV).  The average  E$^*$ of the composite
system $^{204}$Hg computed using {\sc Platypus} is high (42 MeV) due to large
positive Q-value (+12.4 MeV) for $^6$He fusing with $^{198}$Pt. The  major
residue channels  predicted at this  E$^*$ and over the measured range  of
proton  energies are $^{199,200}$Hg.  The $\gamma$-transitions for $^{199}$Hg
are not observed while those from $^{200}$Hg are found  to be populated weakly
(with  the proton gate). Multi-nucleon   transfer   reactions   are   known  to 
 take   place preferentially at  an optimum  Q-value (Q$_{opt}$) obtained  from
the semi-classical   trajectory  matching   condition~\cite{Bro91}.   The
available E$^*$ (Q$_{gg}$-Q$_{opt}$ = 31 MeV) from transfer of $^6$He (Table
\ref{tab3})  is favorable for populating  the residue channels $^{201,202}$Hg,
which  is  consistent  with  the  present  measurement (Fig.~2c).  The same  is 
found to  be applicable  for the  $^5$He~+~$d$ cluster structure of $^7$Li
(breakup  threshold = 9.522 MeV, fusion Q value =  +6.75 MeV). The  average
E$^*$ calculated from  {\sc Platypus} for this   combination  (Table  
\ref{tab3})   favors  residue   channels $^{198,199}$Hg for which the  $\gamma$
transitions are not visible in the $d$ gated spectrum. While the lower E$^*$
estimated from transfer Q-values  is more  suited for  populating $^{200}$Hg, 
in concurrence with the data (Fig~4b). It is worth noting that, although not
included in the present calculations, the unstable $^{5}$He fragment (half-life
of $7.6 \times 10^{-22}$ s) can be dissociated into $\alpha + n$ near the
$^{198}$Pt target. Events where the alpha particle is captured and the neutron
survives along with the deuteron might happen, forming the $^{202}$Hg
compound-nucleus with E$^*$ $\sim 25$ MeV. These events compete with those
involving the $^{7}$Li binary fragmentation into $\alpha + t$, where the
$\alpha$-particle gets captured and the triton escapes. However, the latter
component should be dominant based on the smaller breakup-threshold value, i.e.,
$2.47$ MeV compared to $9.522$ MeV. 

 Based on these observations it can be inferred that,  for the capture of
$^{5,6}$He  from the well-bound cluster configurations of $^7$Li,  the large
value  of the breakup threshold does not favor the process of breakup fusion,
unlike  that for the  $t$ and $\alpha$ particles that are  weakly bound in $^7$Li, and massive
transfer from  bound states could be the main process.
  
\begin{figure}[h]
\includegraphics[width=18pc]{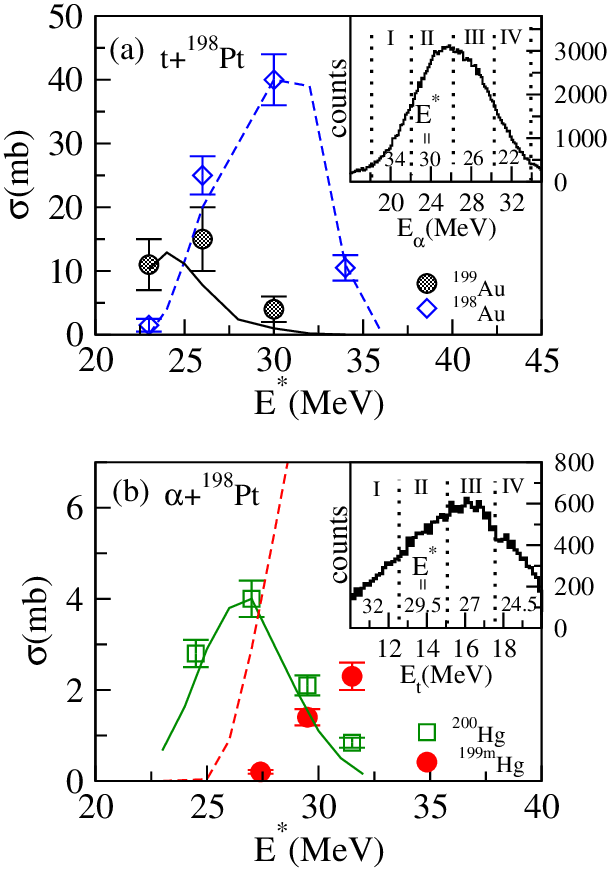}
\caption{(color online) (a) Residue cross-sections  as a function of excitation 
energy (E$^*$) of  the  primary  composite system formed after
$t$-capture.  The E$^*$ in $^{201}$Au, corresponding  to 
kinetic energy  (E$_{\alpha}$)  of the  surviving $\alpha$- particle (shown  in
the inset) is calculated from the classical  trajectory model  of breakup-fusion
-  {\sc platypus}.  (b) same as (a)  but  for $\alpha$-capture. The data points
for $^{199}$Hg (filled circle) represent the lower limit for the cross-sections
for this channel as contributions below the isomeric  state (13/2$^+$) are not
included. The  solid  and  dashed curves  are statistical model calculations
({\sc  pace2}) for two and three neutron evaporation channels, using the angular
momentum distribution and excitation energy of the  primary  composite 
system obtained from {\sc  platypus} (see text). The residue cross-sections of (a) and (b), when summed over E*, are equal to the integrated cross-section at a beam energy of 45~MeV as plotted in Fig. 3a for $t$ and $\alpha$ - capture respectively.}
\label{fig:fig4}
\end{figure}

\begin{table}[htb]
\begin{center}
\caption {Ground state $Q$-values for the $\alpha$, {\it t}, $^5$He and
  $^6$He transfer  channels observed in the $^7$Li  + $^{198}$Pt along
  with  optimum Q  values estimated  from the  semi-classical transfer
  trajectory matching condition. The excitation energies for breakup
  fusion from {\sc platypus} calculation are also tabulated.}
 \begin{tabular}{@{}lllll}
\hline
  exit channel &  $Q_{gg}$  &  $Q_{opt}$    
&   E$^*$   \\
 & & & {\sc platypus} \\
 & (MeV) & (MeV) & (MeV) \\
\hline
  $\alpha$ + $^{201}$Au &  +8.9  & -14.1 & 28.0  \\
  $t$ + $^{202}$Hg &  -2.6   &  -28.6  &  25.5  \\
  $d$  + $^{203}$Hg &   -2.8  &  -28.6  &  38.0 \\
  $p$ + $^{204}$Hg &   +2.4  &  -28.6  &  42.0 \\
\hline 
\end{tabular}
\label{tab3}
\end{center}

\end{table} 

In summary, exclusive measurement of the  outgoing particles and $\gamma$-rays
in conjunction  with  the calculations  made  using a classical dynamical model
have demonstrated  for the first time, the critical role of different cluster
structures of $^7$Li in the dynamics  of reaction mechanism. A  good   agreement
 between   the  calculations  and   the  measured quantities suggests, the
dominant mechanism of capture  of the fragments with low binding
energy in $^7$Li  ($t$ and $\alpha$) after the inelastic  excitation of $^7$Li  above  the  breakup  threshold  is
breakup  followed  by fusion. The low-lying states (up to 4 MeV) in the
continuum populated by the  direct breakup of $^7$Li  make the major 
contribution to the breakup fusion cross-section.   In  case of  capture 
 involving $^5$He+$d$ and  $^6$He+$p$ clusters with relatively high binding energy in $^7$Li,  the evaporation
residues are more  neutron rich  than predicted  from the  model for fusion of 
$^5$He and  $^6$He after the  breakup, suggesting  that the mechanism is not
breakup fusion but could be massive transfer.

 The  cross-section of evaporation  residues for  different excitation energies 
of the  composite system,  formed after  fusion of  $t$ and $\alpha$  particles 
were successfully  explained,  by the  classical dynamical model of
breakup fusion. This information can be useful for studying nuclear structure of
the nuclei formed as $^{5,6}$He +target or  $t$ + target,  using a  $^7$Li beam 
\cite{jung02}.  There  are no existing quantum approaches  for quantifying these
measurements which are planned in  rare isotope beam facilities. Although it is
a great theoretical challenge, the development  of such a quantum approach is
highly desirable. The present  results are useful for the theoretical
developments,  and  also have  implication  in predicting  production
cross-section  of  exotic  nuclei  with  radioactive  ion  beams  for performing
high spin spectroscopy.  It would be interesting to extend such studies with
radioactive nuclei having predominant weakly bound cluster  structures  and to 
develop  the  classical dynamical  model further.

We  acknowledge  the  accelerator  staff  for smooth  running  of  the
machine.  AD-T thanks Leandro Gasques for providing him with Sao Paulo
potential barriers.

\end{document}